\documentclass[12pt]{article}

\usepackage{amsmath,amssymb,amsfonts,amsbsy}
\usepackage{cite}
\usepackage{graphicx}
\usepackage{wrapfig}
\usepackage{epsfig}

\usepackage{bbm} 
\usepackage{bm} 
\usepackage{color}                                                       %
\usepackage{dsfont} 
\usepackage{latexsym} 
\usepackage{lscape} 
\usepackage{mathrsfs} 
\usepackage{morefloats} 
\usepackage{floatflt} 
\usepackage{slashed} 
\usepackage{psfrag}

\DeclareFontFamily{OT1}{mygreek}{}%
\DeclareFontShape{OT1}{mygreek}{m}{n}{<->omsegr}{}%
\DeclareFontShape{OT1}{mygreek}{b}{n}{<->omsegrb}{}%
\DeclareFontShape{OT1}{mygreek}{m}{it}{<->omsegri}{}%
\DeclareFontShape{OT1}{mygreek}{bx}{n}{<->sub * mygreek/b/n}{}%
\DeclareFontShape{OT1}{mygreek}{m}{sl}{<->sub * mygreek/m/it}{}%
\DeclareSymbolFont{Greekrm}{OT1}{mygreek}{m}{n} 
\DeclareSymbolFont{Greekbf}{OT1}{mygreek}{b}{n} 
\DeclareSymbolFont{Greekit}{OT1}{mygreek}{m}{it} 
\DeclareMathSymbol{\omegab}{\mathalpha}{Greekbf}{119}

\begin{document}
\addcontentsline{toc}{subsection}{{Title of the article}\\
{\it Nikolai Nikolaev}}

\setcounter{section}{0}
\setcounter{subsection}{0}
\setcounter{equation}{0}
\setcounter{figure}{0}
\setcounter{footnote}{0}
\setcounter{table}{0}

\begin{center}
\textbf{PRECURSOR EXPERIMENTS TO SEARCH FOR PERMANENT ELECTRIC DIPOLE MOMENTS
  (EDMS) OF PROTONS AND DEUTERONS AT COSY }

\vspace{5mm}

 Andreas Lehrach$^{\,1}$, Bernd Lorentz$^{\,1}$, William
 Morse$^{\,2}$, \underline{Nikolai Nikolaev}$^{\,3\,\dag}$ and \newline
 Frank Rathmann$^{\,1}$
\vspace{5mm}
  
\begin{small} (1) \emph{Institut f\"ur Kernphysik and J\"ulich Center for Hadron Physics,
   Forschungszentrum J\"ulich, 52425 J\"ulich, Germany} \\

  (2) \emph{Physics Department, Brookhaven National Laboratory, Upton, NY
    11973, USA} \\

(3) \emph{Institut f\"ur Kernphysik and J\"ulich Center for Hadron Physics, Forschungszentrum J\"ulich, 52425 J\"ulich, Germany, and Landau Institute, 142432 Chernogolovka, Russia} \\

  $\dag$ \emph{E-mail: n.nikolaev@fz-juelich.de}
\end{small}
\end{center}

\vspace{0.0mm} 

\begin{abstract}
  In this presentation we discuss a number of experiments on the search for
  proton or deuteron EDMs, which could be carried out at COSY-J\"ulich. Most
  promising is the use of an radio-frequency radial electric field flipper
  that would lead to the accumulation of a $CP$ violating in-plane beam
  polarization by tiny spin rotations. Most crucial for storage ring searches
  for EDMs is the spin-coherence time, and we report on analytic evaluations
  which point at a much larger spin-coherence time for deuterons by about a
  factor of 200 compared to the one for protons, and at COSY, the spin
  coherence time for deuterons could amount to about $10^5$ s. \\

\end{abstract}

\vspace{7.2mm} 

\section{Introduction}
Electric dipole moments (EDM) are one of the keys to understand the origin of our Universe. The Universe as we know it has a microscopic net baryon number -- about 0.2 baryons per cubic meter, or $\sim 10^{-10}$ of the density of relic photons. In 1967 Andrei Sakharov formulated three conditions for 
baryogenesis~\cite{BKFSakharov:1967dj}: 
\begin{enumerate}
 \item Early in the evolution of the universe, the baryon number conservation must be violated sufficiently strongly,
 \item the $C$ and $CP$ invariances, and $T$ invariance thereof,  must be violated, and
 \item at the moment when the baryon number is generated, the evolution of the universe must be out of thermal equilibrium.
\end{enumerate}

$CP$ violation in kaon decays is known since 1964, it has been observed in $B$-decays and charmed meson decays. The Standard Model (SM) accommodates $CP$ violation via the phase in the Cabibbo-Kobayashi-Maskawa matrix. $CP$ and $P$ violation entail  non-vanishing $P$ and $T$ violating electric dipole moments (EDMs) of elementary particles $\vec{d} = d\vec{S}$.  Although extremely successful in many aspects, the SM has at least two weaknesses: neutrino oscillations do require extensions of the SM and, most importantly, the SM mechanisms fail miserably in the expected baryogenesis rate. Simultaneously, the SM predicts an exceedingly small electric dipole moment of nucleons  $ 10^ {-33} < d_n < 10^ {-31}$ e$\cdot$cm, way below the current upper bound for the neutron EDM, $d_n <2.9 \times 10^{-26}$ e$\cdot$cm,  and also beyond the reach of future EDM searches \cite{BKFEDMreviews}. In the quest for physics beyond the SM one could follow either the high energy trail or look into new methods which offer very high precision and sensitivity. Supersymmetry is one of the most attractive extensions of the SM and S. Weinberg emphasized in 1992~\cite{BKFWeinberg:1992dq}:  "Endemic in supersymmetric (SUSY) theories are $CP$ violations that go beyond the SM. For this reason it may be that the next exciting thing to come along will be the discovery of a neutron electric dipole moment." The SUSY predictions span typically $10^{-29} < d_n < 10^{-24}$ e$\cdot$cm  and precisely this range is targeted in the new generation of EDM searches~\cite{BKFEDMreviews}. 

There is  consensus among theorists that measuring  the EDM of the proton, deuteron and helion is as important as that of the neutron. Furthermore, it has been argued some 25 years ago that $T$-violating nuclear forces could substantially enhance nuclear EDMs~\cite{BKFFlambaum:1985gv,BKFFlambaum:1985ty}. At the moment, there are no significant {\it direct} upper bounds available on $d_p$ or $d_d$.

Non-vanishing EDMs give rise to the precession of the spin of a particle in an electric field. In the rest frame of a particle 
\begin{equation}
{d\vec{S} \over dt^*} = \mu\vec{S}\times \vec{B}^*+ \vec{d}\times \vec{E}^*,
\label{eq:precession}
\end{equation}
where in terms of the lab frame fields
\begin{eqnarray}
\vec{E}^*= 
\gamma( \vec{E}+\vec{\beta}\times \vec{B})\, ,\nonumber\\
\vec{B}^*= 
\gamma(\vec{B}-\vec{\beta}\times \vec{E})\, .
\label{eq:Lorentz}
\end{eqnarray}
While  ultra-cold  electrically neutral atoms and neutrons can conveniently by
stored in traps, the EDM of charged particle can only be approached with
storage rings \cite{BKFKhriplovich}. EDM searches of charged fundamental 
particles have hitherto been impossible, because  of the absence of the required new class of electrostatic storage rings. An ambitious quest for a measurement of the EDM of the proton with envisioned sensitivity down to $d_p \sim 10^{-29}$ e$\cdot$cm is under development at BNL~\cite{BKFSemertzidis:2009zz}. The principal idea is to store protons with  longitudinal polarization in a purely electrostatic ring: the EDM would cause a precession around the radial electric field and thus lead to a build-up of  transverse polarization which could be measured by standard polarimetry. Related ideas on dedicated storage rings for the deuteron and helion EDM are being discussed at IKP of Forschungszentrum J\"ulich within the newly found JEDI collaboration\footnote{J\"ulich Electric Dipole moment Investigations}. 

Before jumping into construction of dedicated storage rings, it is imperative
to test  technical issues at existing facilities. Here we review several ideas
for precursor experiments which could be performed at COSY subject to very modest additions to the existing machine. In
a magnetic ring like  COSY, the stable polarization axis in the absence of
longitudinal magnetic fields is normal to the ring plane, and at the heart of
the most promising proposal is a radio-frequency electric field (RFE) spin
flipper which would rotate the spin into the ring plane. The resulting  
EDM-generated $P$ and $T$ non-invariant
in-plane polarization which can be determined from the up-down asymmetry of
the scattering of stored particles on the polarimeter. Unless show stoppers
like false spin rotations via the magnetic moment pop up, one could theoretically aim for an upper bound for the deuteron of  $d_d < 10^{-24}$ e$\cdot$cm, which would be as valuable as the existing upper bounds on $d_n$~\cite{BKFFlambaum:1985gv,BKFFlambaum:1985ty}.

\section{EDM searches: state of the art}
The question of whether particles possess permanent electric dipole moments
has a long-standing history, starting from the first search by Smith,
Purcell, and Ramsey~\cite{BKFPhysRev.108.120} for a neutron EDM, which, over the
last 50 years or so, resulted in ever decreasing upper limits. In
Table~\ref{table2}, we give current and anticipated EDM bounds and
sensitivities for nucleons, atoms, and the deuteron and a rough measure of their probing power relative to the neutron ($d_n$). At this level, storage ring EDM measurements bear the potential of an order of magnitude higher sensitivity than the currently planned neutron EDM experiments at SNS (Oak Ridge), ILL (Grenoble-France), and PSI (Villigen, Switzerland)~\cite{BKFlamoreux}.

\begin{table}[hbt] 
\centering
    \begin{tabular}{l|r|l|l|l}
      Particle	& Current  Limit & Goal  & $d_n$ equivalent &reference  \\ \hline
      Neutron	& $<2.9 \times 10^{-26}$  & $\approx 10^{-28}$	& $10^{-28}$                                       & \cite{BKFBaker:2007df}\\
      $^{199}$Hg	& $<3.1 \times 10^{-29}$  & $10^{-29}$  & $10^{-26}$                                       & \cite{BKFGriffith:2009zz}\\
      $^{129}$Xe	& $<6.0 \times 10^{-27}$  & $\approx 10^{-30} - 10^{-33}$  & $\approx 10^{-26} - 10^{-29}$ & \cite{BKFPhysRevLett.86.22}\\   
      Proton	& $<7.9 \times 10^{-25}$  & $\approx 10^{-29}$  & $10^{-29}$                                       & \cite{BKFGriffith:2009zz}\\
      Deuteron  &             & $\approx 10^{-29}$  & $3\times 10^{-29} - 5 \times 10^{-31}$ \\ \hline  
    \end{tabular}
\caption{\footnotesize Current EDM limits in units of [e$\cdot$cm], and long-term goals for the neutron, $^{199}$Hg, $^{129}$Xe, proton, and deuteron are given here. Neutron equivalent values indicate the EDM value for the neutron to provide the same physics reach as the indicated system.}
\label{table2}
\end{table}

\section{Search for electric dipole moments of protons, \newline deuterons, and $^3$He at COSY\label{sec:edm}}

COSY has a history of highly successful operation of cooled polarized beams
and targets -- in fact, COSY is a unique facility for spin physics with
hadronic probes on a world-wide scale. The IKP-COSY environment is 
ideally suited for a major (medium-sized) project involving spin and storage
rings as it will be required for the search for permanent EDMs of charged
fundamental particles ({\it e.g.}, protons, deuterons, and other light
nuclei). JEDI is planning to search for EDMs of the proton and other charged particles in a storage ring with a statistical sensitivity of $\approx 2.5 \times 10^{-29}$ e$\cdot$cm per year, pushing the limits even further and with the potential of an actual particle-EDM discovery.

The proposed new method employs radial electric fields (and magnetic fields)
to steer the particle beam in the ring, electric quadrupole magnets to form a 
weak focusing lattice, and internal polarimeters to probe the particle spin
state as a function of storage time. An RF-cavity and sextupole magnets will
be used to prolong the  spin coherence time (SCT) of the beam. For protons, it
requires building a storage ring with a highly uniform radial E-field with
strength of 
approx. $17$\,MV/m between stainless steel plates about 2\,cm apart. The
bending radius will be approx. $25$\,m, and including the straight sections
such a machine would have a physical radius of approx. $30$\,m.  The so-called magic momentum of 0.7\,GeV/$c$ (232\,MeV), is the one where the $(g-2)$ precession frequency is zero (see Table~\ref{table1}).
\begin{table}[hbt] 
\centering
 \begin{tabular}{c|c|c|c}
    Particle & $p$ (GeV/$c$)  & $E$ (MV/m) & $B$ (T) \\ \hline
    Proton   & 0.701   & 16.8  & 0  \\
    Deuteron & 1.000   & -4.03 & 0.16 \\
    $^3$He   & 1.285   & 17.0 & -0.051 \\ \hline
  \end{tabular}
\caption{\footnotesize Parameters for the transverse electric and magnetic fields required to freeze the spin in an EDM storage ring of radius $r=30$\,m.}
\label{table1}
\end{table}

\section{Precursor experiments at COSY}
The above cited tentative upper bound for the proton EDM as part of a nucleus
in an electrically neutral atom, $|d_p| < 7.9 \times 10^{-25}$\,e\,cm, derives
from the {\it theoretical  reinterpretation} of the upper bound for the EDM of $^{199}$Hg~\cite{BKFGriffith:2009zz}. We briefly review here possible first  {\it direct} measurements of an upper limit for the proton and deuteron EDM using a normal magnetic storage ring like COSY. One needs to isolate a CP-violating  precession of the spin caused by an electric field. Such experiments are widely considered must-do experiments, before embarking on the development and construction of storage rings  with electrostatic deflectors. 

\subsection{RFE spin rotator with Siberian snake}
\begin{figure}[htb]
\begin{center}\begin{minipage}{7.3cm}
   \includegraphics[width=6.5cm,clip]{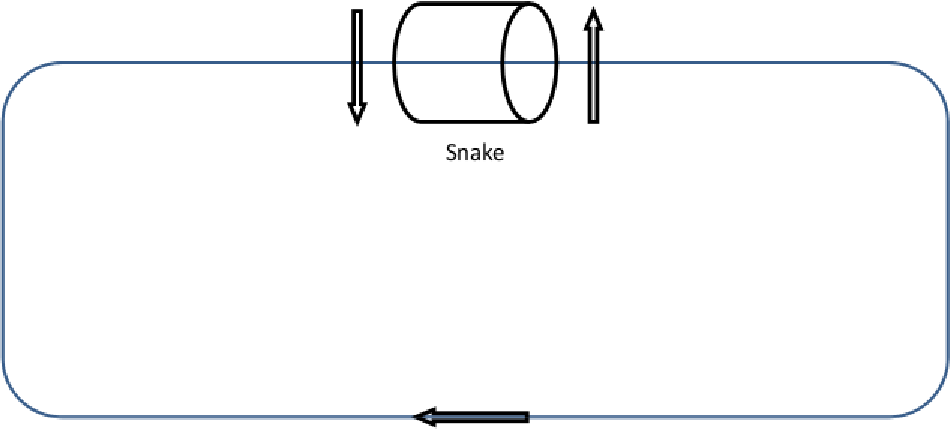}
\vspace{0.3cm}\\
   \includegraphics[width=6.5cm,clip]{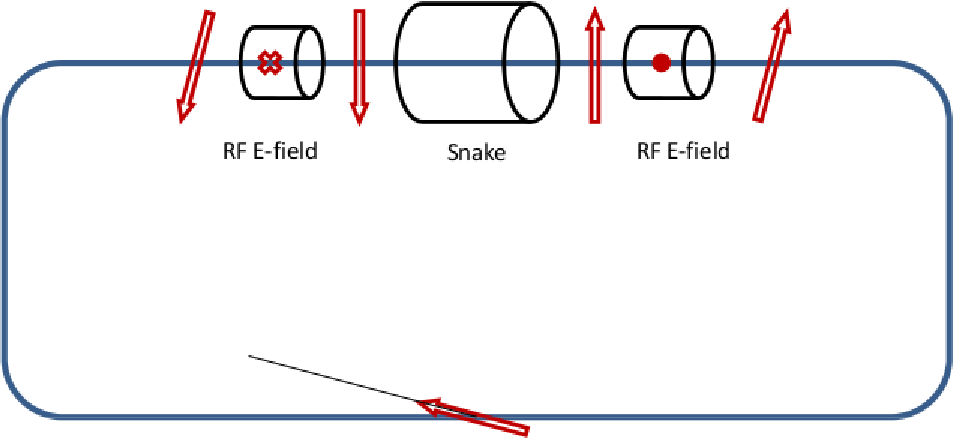}
\vspace{0.3cm}\\
   \includegraphics[width=6.5cm,clip]{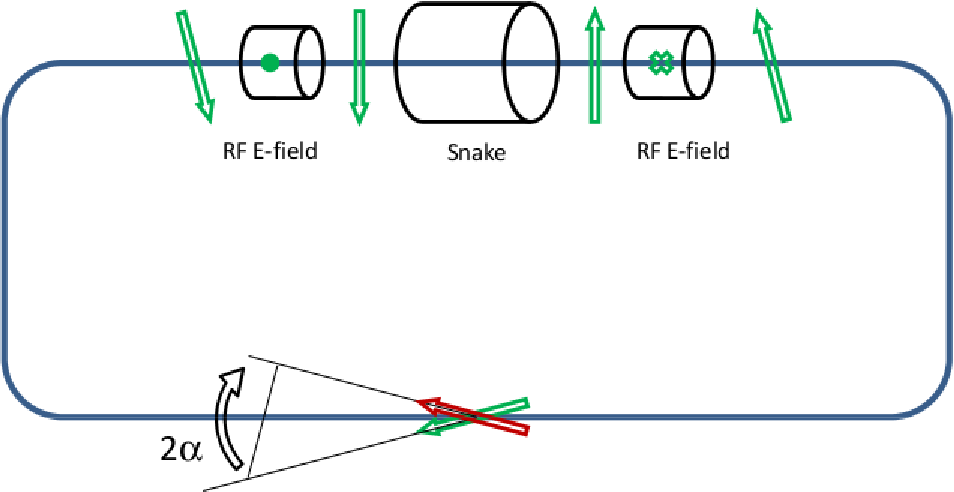} 
\end{minipage}
\begin{minipage}{7.5cm}
   \caption{\label{edmfig}Concept of a proton EDM measurement using a Siberian snake in COSY.  Top panel: Using the snake, the spin closed orbit is aligned along the direction of motion of the proton beam in the straight section opposite the snake. Middle panel: For odd turns in the machine, an electric RF E-field perpendicular to the ring plane in front and behind the snake rotates the stable spin axis by a small angle $\alpha$ away from the longitudinal direction. Bottom panel: For even turns in the machine where the RF E-field is reversed, the spin closed orbit is then rotated by an angle $2\alpha$, leading after $n$ turns to a depolarization of the beam, $P(n)=P_0 \cdot \cos(2\alpha)^n$.}
\end{minipage}
\end{center}
\end{figure}

Making use of a Siberian  snake in COSY yields a stable longitudinal spin-closed orbit in a target section opposite the snake (see Fig.~\ref{edmfig}, top panel). Using two RF E-field systems in front and behind the snake (middle and bottom panels) allows one to provide a certain degree of depolarization in the beam due  to the torque $\vec{d}\times \vec{E}$, where $d$ denotes the proton electric dipole moment. When the RF E-field is reversed in polarity turn by turn, this torque produces a small mismatch between the two stable spin axes, hence the beam depolarizes. While the angle is exceedingly small ($\alpha\approx 10^{-7}$ rad, see Fig.~\ref{edmfig}, bottom panel), the number of turns $n$ in the machine can be made very large ($n\approx 5\cdot 10^{10}$). The sensitivity of this approach is rather limited to values of $d \approx 10^{-17} - 10^{-18}$\,e\,cm, but a measurement would nevertheless constitute a first {\it direct} measurement of an upper limit for the proton EDM. 

\subsection{Dual beam method: protons and deuterons stored simultaneously}
The dual beam method is equivalent to the $g-2$ measurement of the muon EDM
$d_\mu$, reported in ~\cite{BKFPhysRevD.80.052008}. It seems possible to store 
coasting proton and deuteron beams in COSY simultaneously. The way this would
be achieved is by first injecting deuterons from the injector cyclotron into
COSY and accelerating them to highest energy, where the beam lifetime reaches
hundreds of hours. During the deuteron storage time, the injector cyclotron is
tuned for protons, the stored deuterons in COSY are decelerated to the
injection momentum of approx. $300$ MeV/c, electron cooled, and protons are injected. In order to cool both beams, the electron cooler voltage is switched to match the velocities of protons and deuterons for short time periods of about 10~s.

The search for the muon EDM made use of the fact that the magnetic fields
in the $g-2$ experiment were well known, and one was able to relate the
observed additional amount of spin rotation to the muon EDM. In our scenario,
we would compare the spin precession due to the deuteron EDM using the protons
as a means to determine the magnetic properties of the
machine. Experimentally, the task boils down  to the determination of the
invariant spin axes of the simultaneously stored protons and deuterons using a
polarimeter.
 Assuming a value for the proton EDM, derived from the measurement on $^{199}$Hg  (see Table~\ref{table2}), any mismatch of the invariant spin axes for deuterons and protons would be associated to an upper limit deuteron EDM. The sensitivity of this method to $d_d$ would be similar to the one achieved in the $g-2$ determination of $d_\mu$, {\it i.e.}, amount to about  $d_d=10^{-19}$ $e \cdot cm$.

\subsection{Morse-Orlov-Semertzidis resonance method for EDM measurements in storage rings}
This idea for a measurement using an all magnetic ring is described in \cite{BKFPhysRevLett.96.214802}. One would inject sideways polarization into a machine with a vertical invariant spin axis, the EDM produces a growing vertical polarization $P_y$, and  using two sub-beams with different machine tunes that would be independently modulated, allows one to isolate the EDM of the orbiting particles. The sensitivity of this method for protons is estimated to reach $d_p=10^{-29}$ e$\cdot$cm/yr, but because of systematic errors, the idea is  presently no longer pursued at BNL. For COSY, in terms of precursor EDM measurements, this idea is being considered, although detailed evaluations have not been looked into yet.

\subsection{The resonance EDM effect with RFE flipper}
This is our favorite option and below we unfold its features in some  more detail in the following sections.

\section{The resonance RF electric flipper at COSY}

The idea is to supplement a COSY magnetic ring  with a radiofrequency electric flipper (RFE) which runs at a frequency tuned to the spin tune $G\gamma$. Much of the discussion is for deuterons at COSY but there emerges an interesting option also for protons.

\subsection{Tipping the vertical polarization to the CP-violating in-plane  polarization}

Hereafter we focus on pure vertical ring magnetic field $\vec{B}$ and pure radial flipper field $\vec{E}$. An RFE flipper is added in a section where $\vec{B}=0$. A non-vanishing EDM, $\vec{d}= e d \vec{S}$, gives rise to the precession of the spin $\vec{S}$ in an electric field $\vec{E}$ with $\omega_{\mathrm{EDM}}=e d E$. A single pass through the flipper of length $L$ with a radial electric  field $\vec{E}$ would tilt the initial vertical spin $\vec{S} \parallel S_y$, and generate a longitudinal component $S_z = S_y\cdot \alpha$, where $\alpha = d EL/\beta c$. To appreciate the complexity of the task, for a beam of  deuterons with $T=100$ MeV, a RFE flipper of length $L=1$~m, a realistic electric field of $E=15$~kV/cm, and  $d = 10^{-23}$~cm, one finds $\alpha=2.4\cdot 10^{-12}$.

\subsection{The coherent buildup of the EDM effect: single spin problem}
The so generated longitudinal spin would precess  in the magnetic field of the
ring with respect to the momentum vector with frequency $f_S = \gamma G f_R$,
i.e., by an angle $\theta_S =2\pi\gamma G$
per revolution, where $G$ is the anomalous magnetic moment and $f_R$ is the
ring frequency. The tiny EDM spin
rotations  we are after do not disturb this precession.  Compared to the ring
circumference, such a flipper can be treated as a point-like element.
In view of the minuscule  $\alpha$ the change of the magnitude of the in plane polarization, $S_{||} =(S_x^2 + S_z^2)^{1/2}$, per pass is well approximated by $S_{||}(i+1) = S_{||}(i) + S_y\alpha \cos\theta(i)$. Upon  summing over $k$ passes, one obtains
\begin{equation}
S_{||} =S_y\sum _{l=1}^{k}  \alpha \cos(l\theta_S)  \, ,
\end{equation}
which for a static electric field, $\alpha=$const,  would simply oscillate around zero. Evidently, the electric field of the flipper must by modulated in sync with the precession of the spin: $E= E_0 \cos(l \theta_F ) = E_0\cos(\theta_F f_R t) $, {\it i.e.}, $ \alpha =\alpha_E \cos(\theta_F f_R t)$, resulting in the Master Equation
\begin{equation}
S_{||}(t) = S_y\sum _{l=1}^{k}  \alpha_E \cos(l\theta_S)\cos(l\theta_F) =
{1\over 2}\sum _{l=1}^{k} [\cos(l(\theta_S-\theta_F)) +
\cos(l(\theta_S+\theta_F))]ß, .
\label{eq:master3}
\end{equation}
Only the resonance condition 
\begin{equation}
\theta_F = \pm \theta_S 
\label{erq:resonance}
\end{equation}
furnishes the coherent build-up of the EDM signal (hereafter without loss of
generality $\theta_F =\theta_S$)
\begin{equation}
S_{||}(t)= {1\over 2}S_y\alpha_E \nu t . 
\end{equation}
Swapping the harmonic modulation for the rectangular one would enhance the EDM signal by a factor $4/\pi$:
\begin{equation}
S_{||}(t) = S_y\sum _{l=1}^{k}  \alpha_E |\cos(l\theta_S)| = 
{2\over \pi}S_y\alpha_E \nu t . 
\label{eq:master2}
\end{equation}
For the sake of analytic simplicity, we focus here on the harmonic  RFE
flipper, which must run at a frequency $f_F =G\gamma  f_R = f_S$. By a judicious choice of the particle energy one could readily stay away of depolarizing  resonances in the machine.

\subsection{The RFE flipper disturbs the orbit and the spin tune}
The presence in a ring of an RFE flipper with oscillating electric field would
affect both the particle orbit and spin tune. First, the RFE flipper would
generate an oscillating radial momentum $\Delta p_r = eE_0 L\cos(G\gamma f_R
t_i)/\beta c$ per $i$-th pass, which is off-tune with the ring frequency and
betatron frequency. For the above specified RFE filter and 100 MeV deuterons the bending angle is about $\pm 2\cdot 10^{-3}$, well within the machine acceptance of COSY.  Second, the spin precession with respect to the momentum rotation also acquires an oscillating correction
\begin{equation}
\vec{\omega}= -{e \over m}\left[
G\vec{B}- \left(G-{1\over
      \gamma^2-1}\right)\vec{\beta}\times \vec{E}\right]
\label{eq:BMT}
\end{equation}
to $G \gamma $ familiar for a pure magnetic ring, where the electric term combines the changes of the spin precession proper and of the cyclotron frequency. The net effect can be viewed as a frequency modulation of the spin tune, $G\gamma \rightarrow G\gamma [1-y_F\cos(l\theta_S)]$ where for energies of the practical interest 
\begin{equation} 
y_F \approx \left(1-{1\over G(\gamma^2-1)}\right){\beta E L \over 2\pi B R} 
\end{equation} 
is numerically small (here $R$ stands for the ring radius). Then our Master Equation entails only a time-independent weak reduction of the accumulation rate:
\begin{eqnarray}
S_{||}(t) = S_y\sum _{l=1}^{k}  \alpha_E
\cos(l\theta_S)\cos(l\theta_S[1-y_F\cos(l\theta_S)] ) = \nonumber\\
 S_y\sum _{l=1}^{k}  \alpha_E
\cos^2(l\theta_S)\left[(1- {1\over 2}y_F^2\cos^2(l\theta_S)\right] =
{1\over 2} (1-{3\over 8} y_F^2)S_y\alpha_E \nu t \,. 
\end{eqnarray}

The Farley pitch correction \cite{BKFFarley:1973sc} to the spin tune would be important in the practical experiment when the ring is run for a long time, but it does not effect a flow of our principal arguments.   

\subsection{Polarimetry and bunched vs. coasting beam}
Each individual particle for the first time enters the RFE flipper  with a
certain 
$\vec{S}_y(0)$, which remains stable, and a certain in-plane component
$\vec{S}_{xz}(0)$. 
Upon $k$ revolutions,  the overall polarization vector can be decomposed as 
$ \vec{S}(t) = \vec{S}_y(0) + \hat{R}_y(k)\vec{S}_{xz}(0) +\vec{S}_{\|}(t)$, 
where $\hat{R}_y(k)$ is a matrix of  spin rotation around the $y$-axis upon 
$k$ revolutions and $\vec{S}_{\|}(t)$  is the in-plane polarization generated 
by the RFE flipper. Upon averaging over an ensemble $\langle \vec{S}_{xz}(0) 
\rangle =0$, we keep $\vec{S}_y(0)$ for $\langle \vec{S}_y(0)\rangle$. For a 
finite-length bunch and/or coasting beam our earlier derivation holds for a
 particle which enters the flipper at $t=0$. Particles which are behind by a fraction $0<z<1$ of the ring circumference enter the flipper at a different field advanced by time $\Delta t= z/f_R$ and the modified Master Equation reads    
\begin{equation} 
S_{||}(z,t) = S_y \alpha_E\sum_{l=1}^k \cos(l \theta_S) \cos(l\theta_S +
z\theta_S) = {1\over 2} S_y \alpha_E f_R t  \cos(z\theta_S)\, .
\end{equation}
The bunch can be viewed as point-like and its polarization is uniform if the length of the bunch $z_b$ satisfies the condition $z_b\theta_S \ll 1$. 

\begin{figure}[htb] 
\centering
 \includegraphics[width=6.5cm,clip]{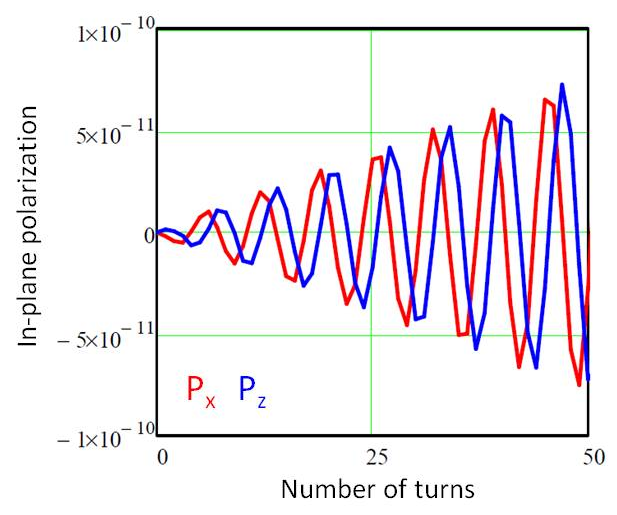}
 \includegraphics[width=6.5cm,clip]{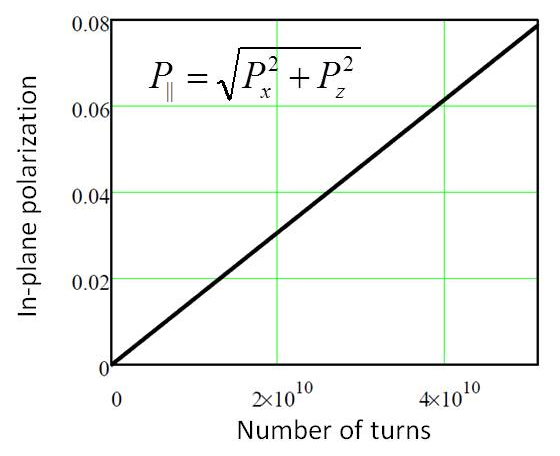}
\caption{Left panel: Oscillating in-plane beam polarization components $P_x$ and $P_z$ ($S_x$ and $S_z$) for the first 50 turns (revolutions) in the machine. Right panel: Evolution of the magnitude of the in-plane polarization $P_\parallel=\sqrt(P_x^2+P_z^2)$ during a spin coherence time of $10^5$~s, which, under the specified conditions using 100 MeV deuterons in COSY corresponds to  a total of $5\times 10^{10}$ turns in  COSY.}
\label{inplane}
\end{figure}
The longitudinal component $S_z$, and the radial one $S_x$ would oscillate, leading to  $S_z=S_{||}(t)\cos (\theta_S f_R t)$ and $S_x= S_{||}(t)\sin (\theta_S f_R t)$, as shown in Fig.~\ref{inplane},  where we show the results of modeling with spin rotation matrices. One would readily extract $ S_{||}(t)$ from the relevant Fourier component of the up-down asymmetry
\begin{equation}
A_{u/d} ={ \int dt [N_{up}(t) - N_{down}(t)]\sin(\theta_S f_R t) \over  
\int dt [N_{up}(t) + N_{down}(t)]} \propto A_N S_{||}(t)\, ,
\end{equation} 
where $N_{up/down}(t)$ are the corresponding count rates --- this is a familiar technique.
A simultaneous measurement of both $S_x$ and $S_z$ would have been an important cross check, but a word of caution is in order: the in-plane magnetic fields are forbidden as the ordinary magnetic moment would cause a false precession of the vertical-to-in-plane spin. This seems to preclude the $S_z$ polarimetry on a longitudinally polarized internal target with longitudinal holding  magnetic field.

The use of a transversely polarized internal target, however, seems  possible, although during the spin-flip process, it must run with empty cell but vertical guide field  switched on, one would inject polarized particles into the cell only at the polarimetry stage, and, since one does not want to change the holding field polarity during the measurement, injection of different hyperfine states from the polarized source is necessary.

\subsection{A null experiment}

Complementing the radial electric field of the RFE flipper with the in-phase 
vertical magnetic field one can realize an exact cancellation of the flipper 
$E$ field by the motional electric field, $\vec{E}^*=0$, see Eq.~(\ref{eq:Lorentz}). This would provide a null experiment for separation  of the genuine EDM signal from false effects. On the other hand, an oscillating motional $\vec{B}^*$ only causes a weak frequency modulation of the spin tune and imposition of $\vec{B}^*=0$ in the EDM run does not seem imperative.

\subsection{Diffusion of the in-plane spin and spin coherence time}

The extremely small single-pass rotation $\alpha_0$ in the RFE flipper can
only 
be overcome by an extremely large number of turns $f_R t$. While the vertical
polarization is preserved by the holding field of the ring, the in-plane spins 
accumulated during the flipper process must all rotate coherently at one and 
the same rotation angle $\theta= \theta_S f_R t$ rather than evolving into 
a hedgehog. The spin coherency is
one of the highest risk factors in all the EDM projects~\cite{BKFSemertzidis:2009zz}. There is an important 
distinction between the lifetime of the polarization along the 
stable-spin axis, the spin coherence time (SCT) of the in-plane 
polarization  when the beam idly rotates in the storage ring and the SCT
during  
the build-up of the in-plane polarization. 

\subsubsection{Spin coherence time for an idle rotation} 
 To a first approximation $S_y$ is preserved irrespective of what happens to 
the rotating in-plane component of the spin. The spin tune $\theta_S =2\pi
\gamma G$ 
varies from revolution to revolution and from stored particle to particle 
because of the momentum fluctuations, $\theta =\theta_S + 2\pi G \delta\gamma  
= \theta_S + \delta\theta$. Hereafter $\theta_S=\gamma_0 G$ and $\gamma_0$ 
is defined for the average beam 
momentum ensured by cooling and RF bucket and by the very definition $\langle
\delta\gamma\rangle =0$. For the beginners, we swamp all imperfections,
nonlinearities, betatron oscillations and
whatever else into a Black Box which generates $\delta\gamma$ on the
turn-by-turn basis, in the future all these effects need to be studied in
detail. The average in-plane spin 
$\langle S_\parallel \rangle$ points at an angle 
$\theta= \theta_S f_R t$, while for an individual particle there is a cumulant
spin precession 
slip $\Delta(k) = \sum_1^k\delta\theta_l $, so that 
\begin{eqnarray}
\langle S_\parallel \rangle =  S_\parallel (0)\langle \cos \Delta \rangle
=S_\parallel (0)\left\{1-{1\over 2}\langle\Delta^2\rangle\right\} = \nonumber\\
=S_\parallel (0)\left\{1-2\pi^2 G^2f_R t  \langle
\delta\gamma^2\rangle\right\} = S_\parallel (0)(1 - t/\tau_{SC,NF})\, ,
\label{eq:Idle1}
\end{eqnarray}
where the subscript NF stands for No Flipper.
It decreases with time because of the angular random walk (diffusion), which 
eventually shall evolve the spin arrow into a hedgehog, and
\begin{equation}
\tau_{SC,NF} \approx   
{ 1\over 2 \pi^2f_R G^2\gamma^2\beta^4 }\cdot \left\langle \left(
\delta p \over p \right)^2\right\rangle^{-1}
\label{eq:Idle2}
\end{equation}
has a meaning of the SCT for an idle rotation in the absence of a spin
flipper.

Admittedly, such a violent turn-by-turn randomization of the momentum 
fluctuations leads to an excessive spin diffusion and arguably 
Eq.~(\ref{eq:Idle2}) gives a lower bound on SCT. A discussion of more realstic scenarios with
slow variations of the beam particle momenta will be reported elsewhere.

\subsubsection{Spin coherence time with a running spin flipper}
Still another source of spin decoherence is the fluctuation 
of the revolution (transit) time $\tau$, described in terms of slip-factor, $\delta \tau/\tau = \eta \delta \gamma/ \gamma \beta^2$, where
\begin{equation}
 \eta = {1\over \gamma_{tr}^2} -{1\over \gamma^2}\,,
\end{equation}
and $ \gamma_{tr}$ is the transition gamma-factor \cite{BKFChao:1999qt}. It
produces a slip of the phase of the RFE flipper per pass 
$ \delta \theta = 2\pi f_F \delta \tau $. Then the Master Equation will take the form  
\begin{equation}
S_{||} =  S_y\alpha_E \sum _{l=1}^{k}\cos(l \theta_S  +\Delta(l))\
\cos(l\theta_S + \eta\Delta(l)i/\beta^2)\, ,
\end{equation}
where $\Delta_i= \sum_{n=1}^{i}\delta\theta_n $ is the cumulant precession
slip  before the $i$-th pass through the RFE flipper.  Following the
derivation of Eq.~(\ref{eq:Idle1}), we readily find
\begin{equation}
S_{||} =  S_y\alpha_E \sum _{l=1}^{k}\cos^2(i \theta_S )(1- l 2\pi^2G^2 C_{SD}^2\langle
\delta^2\gamma\rangle )=S_y\alpha_E
{1\over 2} f_R t \left(1- \pi^2 f_R t  G^2 C_{SD}^2\langle
\delta\gamma^2\rangle \right)\, ,
\label{eq:masterSD}
\end{equation}
where 
\begin{equation}
C_{SD}=1-{\eta\over \beta^2}\, .
\end{equation}

The corresponding SCT equals
\begin{equation}
\tau_{SC} = {2 \over  C_{SD}^2}\tau_{SC,NF} \approx   
{ 1\over C_{SD}^2\pi^2 f_R G^2\gamma^2\beta^4 }\cdot \left\langle \left(
\delta p \over p \right)^2\right\rangle^{-1} \, .
\label{eq:SCT}
\end{equation}
Small $G_d= -0.143$ strongly enhances the deuteron SCT compared to the 
proton SCT (we ignore here a possible difference of $C_{SD}$ for protons and deuterons),
\begin{equation}
\tau_{SC}^p \sim \tau_{SC}^d\cdot \left(\frac{G_d}{G_p}\right)^2\sim
{1\over 200}\tau_{SC}^d\, .
\label{eq:protonSCT}
\end{equation}
  
For non-relativistic particles $-\eta/\beta^2 \approx 1/\beta^2$ and the 
in-plane spin-diffusion  is entirely dominated by the flipper phase slip and
large $C_{SD}^2$ strongly suppresses $\tau_{SC}$. There are two strategies: either find 
a way to suppress $C_{SD}$ or eliminate the flipper phase slip, {\it i.e.}, 
enforce $\eta=0$. Strongly different $G_p$ and $G_d$ suggest the former 
strategy for protons and the latter for deuterons. 

The SCT considerations do obviously favor running COSY at non-relativistic
energies. 
For the reference case of 100 MeV deuterons, $\nu \approx 511$ kHz, and 
cooled  beam with $\delta p / p = 10^{-4}$, our estimate yields 
$\tau_{SC}^d(\eta=0) \sim 3\cdot 10^5$\, s. 

A  purely  electrostatic rings would share the above spin 
decoherence mechanisms, although the analytic treatment would 
be substantially different from that for the point-like RFE flipper.

\subsection{Running RFE flipper at higher frequencies?}

\subsubsection{Bad news for nonrelativistic deuterons?}
Short bunches offer the possibility of operating the RFE flipper  at higher
frequency. One could run the flipper at any frequency $f_F = (\gamma G+K)f_R$,
where $K =0, \pm 1, \pm 2... $ is integer.  Indeed, short bunches probe the $E$-field only at discrete times $t_i = i/f_R$ and $\cos(2\pi l f_F) = \cos(l\theta_S + 2\pi l K)= \cos(l\theta_S)$. Evidently, for an ideal particle, the  build-up of the EDM signal wouldn't depend on $K$. The limitation on the bunch length becomes much more stringent, though: 
\begin{equation}
z_b (\theta_S + 2\pi K)= x_b\theta_S\left(1+ {K\over \gamma G}\right)  \ll 1\,
.
\end{equation}
A similar bound is imposed on the length of the flipper, $z_F$, in units of the ring circumference: $z_F \theta_S (1+ K/\gamma G) \ll 1$.

Simultaneously, the troublesome flipper phase slip acquires the same factor
$(1+ K/\gamma G)$, 
so that $C_{SD}$ in the diffusion rate will change to
\begin{equation}
C_{SD}=1-{\left(1+{K\over \gamma G}\right)\cdot {\eta\over
      \beta^2}} \, .
\label{eq:magic}
\end{equation} 
For deuterons at COSY, $K/|G_d| \gg 1$ and running at higher frequencies invites  an unwanted suppression of the SCT for nonrelativistic deuterons by still another small factor $\sim (G_d/K)^2$, {\it i.e.}, by almost two orders in magnitude.

\subsubsection{Good news for protons: suppression of spin diffusion at magic energies}

A closer look at Eq.~(\ref{eq:magic}) suggests an intriguing possibility of a set of magic energies at which the flipper phase slip would compensate the effect of the spin tune slip. We recall that $\eta$  is large and negative valued for non-relativistic particles. Then by a judicious choice of $K=-N$ and $\gamma$ such that
\begin{equation}
K+\gamma G <0
\end{equation}
one could arrange for $C_{SD}=0$, {\it i.e.,} for a vanishing spin diffusion rate. These magic energies are roots of an equation
\begin{equation}
\gamma^3 = -{K\over G} + 
{\gamma^3 \over \gamma_{tr}^2}\left({K\over \gamma G_p}+1\right)\, .
\label{eq:magic2}
\end{equation}  
For protons $G_p=1.793$ and solutions do exist for $-K=N=2,3,...$. Because the
transition energy is high, $\gamma_{tr}^2 >> 1$ (in one of regimes at COSY  $\gamma_{tr}^2 \approx 3.3$), for a quick estimate of lowest roots one can resort to an iterative solution
\begin{equation}
\gamma_{N-1}= \left(N\over G_p\right)^{1/3}\left(1-
{1\over 3\gamma_{tr}^2}\left[\left(N\over G_p\right)^{2/3}-1\right]\right)\, .
\end{equation}
With the above specified $\gamma_{tr}$, the lowest magic energy at $N=2$ equals $T_p \approx 29$~MeV, which is too
low.  
The second root at $N=3$ corresponds to  $T_p \approx  133$~MeV, which is
within the range of the existing COSY electron cooler. Besides a possibility
of
cooling, this magic energy is preferred because of longer beam lifetime. The third root gives  $T_p \approx 210$~MeV. An asymptotic  convergence of large-$N$ magic energies to transition energy, {\it i.e.}, to an isochronous ring, is noteworthy:
\begin{equation}
\gamma_N^2 =\gamma_{tr}^2 - \beta_{tr}^2 \gamma_{tr}^5 G_p \cdot{1\over N}\, .
\end{equation}
This finding of spin-decoherence-free magic energies lifts the pessimism of Eq.~(\ref{eq:protonSCT}) and paves the way to a 
high sensitivity searches for the proton EDM at COSY.

We strongly emphasize that the existence of magic energies only depends on the
fact that the spin precession and flipper phase slips are locked to each
other and does not depend on the exact model for the phase slip and for the
momentum fluctuations.

\subsubsection{Magic energies for deuterons at COSY}

Deuterons also possess a sequence of magic energies albeit at higher
energies. Since $G_d <0$, here we look for $ K=+1,2,\dots$. To a first
approximation, deuterons and protons do share the same $\gamma_{tr}$. 
Assuming above $\gamma_{tr}$, the lowest magic energy at $K=1$  equals $T_d\approx 0.9$~GeV, while at $K=2$ our estimate  is $T_d\approx 1.15$~GeV, which are accessible at COSY. Transition energy is tunable, for instance at $\gamma_{tr}^2=4$ we find the deuteron magic energies $T_{d}(K=1)\approx 1.03$~GeV and $ T_{d}(K=1)\approx 1.33$~GeV.

Magic energies vindicate the harmonic modulated RFE flippers but leave  open
an issue of dynamic magnetic fields generated by $dE/dt$. For deuterons this
menace, which deserves a separate treatment, can be circumvented by a flat-top
RFE flipper.

\subsection{Advantages of a rectangular (flat-top)  modulated RFE flipper for deuterons}
 
\subsubsection{Even mode flat-top flipper}

If running COSY with deuterons at magic energy would prove impractical, then the phase-slip of the flipper $E$-field emerges as a potential show-stopper for deuterons. Here we notice that this phase-slip can be entirely  eliminated by employing a rectangular (flat-top) modulation of the RFE flipper,
\begin{eqnarray}
 E(t) = E_0(-1)^{N_F}\, , ~~~~~ N_F= \mathrm{int}(\gamma G f_R t/\pi)\, .
\end{eqnarray}
which does not depend on  the phase slip. The exact rectangular modulation is not imperative, what we are asking for is a flat top when the bunch passes through the flipper. In order to avoid the effects of dynamical magnetic fields generated by  $dE/dt$, the $E$-field must be  inverted when the bunch is at 180 degree, the opposite side of the ring. The simplest solution is to lock the  RFE flipper frequency to the ring frequency
\begin{equation}
f_F = {1\over 2N}f_R\, .
\end{equation} 
For deuterons $N=3$, {\it i.e.}, $\gamma |G_d| = 1/2N=1/6$ is a convenient option: here the flipper field is inverted once per $N=3$ revolutions of the beam. However, that demands for somewhat higher kinetic energy: $\gamma_d=1.169$, $T_d=317$~MeV, $\beta=0.52$, $f_R=0.98$~MHz, and $\nu_F=163$~kHz.  The price tag for the higher energy of deuterons is a somewhat shorter spin coherence time: our Eq.~(\ref{eq:SCT}) for $\eta=0$ gives 
$\tau_{SC}\sim 6\cdot 10^3$\, s.

\subsubsection{Odd-mode flat-top flipper: dedicated low-energy ring for the deuteron EDM?}

Curiously enough, for the reason that $1/|G_d| = 7.0145$, the condition $\gamma |G_d| =1/7$ is met at $\gamma= 1.00207$, {\it i.e.}, $T_d= 3.88$ MeV. Such deuterons will make 7 revolutions and pass the flipper 7 times per single spin turn. Then the flat-top cycle can be organized as follows: 

Switch the flipper on when the bunch is on the opposite side of the
ring. After 3 revolutions at $E >0$ switch the field off so that the 4-th
revolution is at $E=0$, and then switch the flipper on again  at inverted polarity, $E<0$, when the bunch is opposite the flipper. The second inversion of the $E$-field is after the 7th revolution with a bunch opposite the flipper. This way we managed to exclude the 4th revolution which would have crossed a flipper at exactly the time when the $E$-field is inverted. 

Low energy enhances both the single-pass tilt of the spin and spin coherence time but decreases the beam lifetime --- the latter might prove a show stopper. Whether one can gain or not in sensitivity to EDM with such a curious option is worth of further scrutiny. 

\subsubsection{Half-integer-mode flat-top flipper}

Still another interesting option is $1/\gamma |G_d| = 6.5$, when
$\gamma=1.07915$ and $T_d=148.5$~MeV. The flipper period would comprise two
spin turns and 13 revolutions of the beam and the sought for cycle must be organized as follows:

The flipper field $E>0$ is switched on when the bunch is at 180 degree from the flipper, kept constant for revolutions 1, 2 and 3, inverted to $E<0$  for  revolutions 4, 5 and 6,  switched off, $E=0$, during the 7-th revolution, inverted to $E>0$ for revolutions 8, 9 and 10,  and $E<0$ for revolutions 11, 12 and 13. 

Running the flipper in such a mode is a challenging task, but an obvious benefit is the smaller $T_d$ and the larger spin coherence time of $\tau_{SC}\sim 3\cdot10^4$s.    
 
\subsubsection{One-third-integer-mode flat-top flipper}  
 
A still more interesting option is $1/|G_d| = 20/3$, when the flipper period comprises 3 spin turns and 20 revolutions of the beam. The flipper field inversion pattern is as follows: $E>0$ for revolutions 1, 2 and 3;  $E<0$ for revolutions 4, 5, 6 and 7; $E>0$  for revolutions 8, 9 and 10; $E<0$ for revolutions 11, 12 and 13 $E>0$ for revolutions 14, 15, 16 and 17, and $E<0$ for revolutions 18, 19 and 20. In this mode $\gamma=1.0522$ and $T_d=98$~MeV, and as we evaluated above,  $\tau_{SC}\sim  10^5$s.    

\section{RF magnetic flipper}
\subsection{A proof of the principle at COSY}
Remarkably, much of the spin dynamics in the suggested EDM experiment  at COSY
can be tested by swapping the RF electric flipper for an RF magnetic flipper
(RFB) with a {\it radial} RF magnetic field. In such an RFB flipper the
magnetic moment of the deuteron would do exactly the same job as the sought
for EDM in the RFE flipper. The anomalous magnetic moment of the  deuteron is
$\sim 3\cdot 10^{-15}$\,e\, cm, while we speak of EDM of $\sim 10^{-24}$\,e\,cm,
consequently a single-pass magnetic tilt $\alpha_B$ can be made 
gigantic compared to
the above estimated $\alpha_E$ for the expected EDM. This adds an entirely  new
dimension: while we dream of accumulation of a several per mill to several
 per cent in-plane
polarization running RFE flipper for $10^5$s, employing an  RFB one could
readily have single-pass tip angles $\alpha_B \sim 10^{-6}$. The net result
will be that within seconds the accumulation of the in-plane polarization will
end up in total consumption of the initial vertical polarization $S_y(0)=+1$,
{\it i.e.}, ideally  we get $S_\parallel=1$ at  $S_y=0$, which then will be
followed by the accumulation of the vertical polarization from the in-plane
one down to $S_y=-1$ at $S_\parallel=0$ and so forth. If the in-plane spin
decoheres, the restoration of the vertical spin will be imperfect and the
decay time of oscillations can be related to the spin coherence time. 

However, an RFB flipper with a {\it longitudinal} magnetic field, 
tangential to the orbit, is doing exactly the same job! Indeed, the 
RFB flipper with {\it radial} field generates resonance forward and 
backward  tips of the spin, which then precesses in the ring magnetic 
field.. The effects of the {\it longitudinal} vs. {\it radial} $B$-fields 
only differ by swapping 
$S_z$ and $S_y$, {\it i.e.}, by a $\pi/2$ shift of the spin precession 
angle, otherwise the buildup of the in-plane polarization is exactly the same.

Remarkably, such a proof of  principle  with {\it longitudinal} RFB flipper
has already been achieved at COSY in January 2011, the analysis is in
progress  and preliminary results have
been reported at several meetings~\cite{BKFStephenson,BKFGuidoboni}.  The period of
oscillations is obviously $\propto 1/\alpha_B$, {\it i.e.}, 
inversely proportional to RFB flipper magnetic field, 
which has indeed been seen in the COSY experiment~\cite{BKFStephenson,BKFGuidoboni}.

\subsection{Systematics and ring imperfections with RFB  flipper}
The beauty of the COSY experiment with gigantic $\alpha_B$ is 
that one could have resorted to a
conventional polarimetry of the oscillating vertical polarization $S_y$. 
In the EDM experiments with $S_\parallel$ in at most per mill range a  
variation of $S_y$ can not be detected, which makes mandatory the 
polarimetry of the precessing in-plane polarizations $S_x$ and $S_z$. 
Various sequels to the COSY experiment could distinguish the spin decoherence 
caused by RFE and RFB  flippers and the one from the ring
imperfections. 

The former has been our major concern, the latter is for the most part an
uncharted territory. We notice that running at magic energy one would
eliminate the flipper effects and the remaining spin decoherence is a
direct measure of the systematic effects driven by the ring imperfections.
A second option has already been tried at COSY \cite{BKFStephenson,BKFGuidoboni}:
rotate the vertical polarization to pure horizonal one, let $S_\parallel$
precess for a long time and rotate it back to the vertical one. This requires
a perfect timing when the RFB flipper is turned on again: as we discussed
in Section 5.4., a slip of the flipper phase by  $\theta_{slip}$ with respect 
to spin precession could suppress the recovered vertical polarization 
$\propto cos(\theta_{slip})$, which would imitate a spin decoherence. Much 
more advantageous is to look at a decay of oscillating $S_x$, 
which would measure spin decoherence in idle precession and give
$\tau_{SC,NF}$ compounded by the possible decoherence from the ring imperfections. 

The RFB flipper of the COSY experiment
 was run in a harmonic mode at a frequency $f_F= (1+\gamma |G_d|)f_R$. 
For a better insight into spin decoherence mechanisms one needs to 
repeat the experiment at lower frequency $f_F = \gamma G_d f_R$ and 
test the predicted suppression of decoherence with flat-top modulated RFB. 
The experiments with protons are equally important to test the predicted 
change of $\tau_{SC}$ from deuterons to protons and to test the predicted
existence of magic energies, as well as a search for a predicted magic 
energy of deuterons in the vicinity of $T_d \sim 1 $~GeV. At last but not 
the least, decreasing the RFB field from micro- to nano- to
pico-tesla range one could explore the systematics of the COSY ring down 
to the anticipated sensitivity of 
the EDM-experiments at COSY.

\section{Summary and Conclusions}

We reported on the first look into the potential of all-magnetic rings as EDM machines. The emerging strategy of the proton and deuteron EDM searches at COSY is as follows:

Running the COSY ring, supplemented with the above specified $20/3$-mode flattop-modulated RFE flipper (for 98 MeV deuterons, $\nu_F = \gamma|G_d|\nu \sim 77$ kHz), for $\tau_{SC}^d = 10^{5}$ and assuming $d_d = 10^{-23}$ e$\cdot$cm, the accumulated $CP$ violating in-plane polarization of the deuteron could be as large as $S_{||} = 0.08 $. To reach an upper bound of $d_d = 10^{-24}$ e$\cdot$cm polarizations of $S_{||}=0.008$ need to be determined, which is within the reach of state of the art polarimetry. Such an upper bound on the deuteron EDM of $d_d < 10^{-24}$ e$\cdot$cm would be comparable to the results from the model-dependent reinterpretation of upper bounds on atomic EDMs~\cite{BKFEDMreviews}, and size-wise is close to the ball-park neutron EDM bounds. 

Magic energies at which the in-plane spin decoherence is strongly suppressed
open entirely new perspectives for the proton (and perhaps the deuteron) EDM
at COSY. The existence of magic energies is a model-independent feature of the 
rotation of the spin by a radiofrequency flipper.  

True, regarding the systematics, we have presently touched only the tip of the
iceberg in a very crude analytic approach and much more scrutiny 
of the ring lattice and imperfections which will affect polarization lifetime
and also somewhat limit the sensitivity is in order.
Specifically, one badly needs spin tracking tools capable of handling with
controlled precision up to $\sim 10^{11}$ turns in a realistically modeled
machine.
A special care must be taken of false rotations via the magnetic moment in
the RFE flipper - these might prove a main systematics and 
has to be thoroughly investigated.
 With all  reservations, the RFE flipper experiment at COSY looks like a promising one. We especially emphasize again here  the importance of extending further in-situ studies at COSY using very slow RF magnetic flippers to study systematic effects for both deuterons and protons. In the case of protons a confirmation of the existence of magic energies, and how well the spin decoherence is eliminated at these energies, need to be studied in dedicated RF magnetic flipper experiments --- this chance of making COSY the proton EDM machine need not be overlooked.
\section{Acknowledgments}
We acknowledge fruitful discussions with Rudolf Maier, Yuri
Senichev, Ed Stephenson, Hans  Str\"oher and Richard Talman. We are grateful to the Workshop
organizers for a chance to present our new ideas on the search for EDM at pure
magnetic rings.


\end{document}